\documentclass[aps,prl,10pt,preprintnumbers,amsmath,amssymb,twocolumn,superscriptaddress,floatfix]{revtex4-2}

\usepackage{amssymb}
\usepackage{amsmath}
\usepackage{epsfig}
\usepackage{hyperref}
\usepackage{xcolor}
\usepackage{slashed}
\usepackage{dsfont,bbm}
\usepackage{mathrsfs}
\usepackage{extarrows}

\usepackage{multirow, graphicx, placeins}

\graphicspath{{./figs/}}

\begin{document}

\preprint{USTC-ICTS/PCFT-26-92}
\title{Full-Color Two-loop Four-point Form Factor in \texorpdfstring{${\cal N}$=4}{N=4} SYM with Global CK-duality}
\author{Zhiming Cai}
\email{caizhiming@itp.ac.cn}
\affiliation{Institute of Theoretical Physics, Chinese Academy of Sciences, Beijing 100190, China}
\affiliation{School of Physical Sciences, University of Chinese Academy of Sciences, Beijing 100049, China}
\author{Yuanhong Guo}
\email{guoyuanhong@imada.sdu.dk}
\affiliation{Center for Quantum Mathematics, Department of Mathematics and Computer Science,
University of Southern Denmark, Campusvej 55, 5230 Odense M, Denmark}
\author{Gang Yang}
\email{yangg@itp.ac.cn}
\affiliation{Institute of Theoretical Physics, Chinese Academy of Sciences, Beijing 100190, China}
\affiliation{School of Physical Sciences, University of Chinese Academy of Sciences, Beijing 100049, China}
\affiliation{School of Fundamental Physics and Mathematical Sciences, Hangzhou Institute for Advanced Study, UCAS, Hangzhou 310024, China}
\affiliation{Peng Huanwu Center for Fundamental Theory, Hefei, Anhui 230026, China}

\begin{abstract}
We determine the full-color two-loop four-point MHV form factor of the stress-tensor supermultiplet in $\mathcal{N}=4$ super-Yang-Mills theory. Using color-kinematics (CK) duality and unitarity cuts, we construct complete one- and two-loop integrands manifesting global off-shell CK duality without explicit $\mu^2$ dependence. Integration-by-parts reduction to uniform-transcendentality master integrals yields purely rational numbers multiplying simple kinematic prefactors, manifesting maximal transcendentality. Expressing the form factor in terms of pentagon functions, we verify full-color infrared-pole cancellation, extract the finite remainder, and obtain new non-planar results for the two-loop triple-collinear splitting function and double-soft current.
\end{abstract}

\maketitle 

\section{Introduction}

Multi-loop calculations beyond the planar limit for processes with high multiplicity is a central frontier in gauge theory and collider physics. In $\mathcal{N}=4$ super-Yang-Mills (SYM) theory, form factors of the half-BPS stress-tensor supermultiplet $\mathcal{T}$ serve as an ideal theoretical laboratory bridging on-shell amplitudes, periodic Wilson loops, and QCD Higgs-plus-multijet amplitudes. In this Letter, we compute the full-color two-loop four-point maximally-helicity-violating (MHV) form factor,
\begin{equation}
\mathbf{F}_{4} = \int d^D x\, e^{-i q\cdot x} \langle p_1 p_2 p_3 p_4|\mathcal{T}(x)|0\rangle \,,
\end{equation}
where $p_i^2=0$ ($i=1,\dots,4$) and $q=\sum_{i=1}^4 p_i$ is the off-shell momentum carried by $\mathcal{T}$.

Notable progress has recently been achieved at two loops for full-color five-point massless scattering in both $\mathcal{N}=4$ SYM~\cite{Abreu:2018aqd,Chicherin:2018yne} and QCD~\cite{Agarwal:2023suw,DeLaurentis:2023nss,DeLaurentis:2023izi}, as well as for the planar four-point form factor~\cite{Guo:2022qgv,Dixon:2022xqh,Guo:2024bsd,Dixon:2024yvq,Li:2024rkq,He:2026ddp,He:2026kdq} and the leading-color Higgs-plus-four-parton amplitude in the heavy-top limit~\cite{DeLaurentis:2026brm}. While color-kinematics (CK) duality~\cite{Bern:2008qj,Bern:2010ue} (see \cite{Bern:2019prr,Bern:2022wqg,Adamo:2022dcm} for reviews) has enabled full-color constructions up to five and four loops for two- and three-point form factors~\cite{Boels:2012ew,Yang:2016ear,Lin:2021kht,Lin:2021lqo}, extending both the CK-dual integrand and its analytic evaluation to the full-color two-loop four-point form factor has remained an open challenge. Unlike lower-point cases, the four-point kinematics spans four dimensions, introducing permutation-dependent tree factors and subtle $D$-dimensional loop-momentum terms.

Here we obtain the complete full-color two-loop four-point form factor from integrand to function level. Using CK duality and unitarity cuts~\cite{Bern:1994zx,Britto:2004nc}, we construct one- and two-loop integrands that manifest global off-shell CK duality. A key ingredient is a spinor-helicity prefactor $\beta_{1234}$ carrying a Gram-determinant pole, which enforces minimal UV power counting and captures $D$-dimensional information without any explicit $\mu^2$ dependence—in contrast to the planar integrand of~\cite{Guo:2024bsd}. Integration-by-parts reduction onto five-point one-mass uniform-transcendentality (UT) master integrals~\cite{Abreu:2020jxa,Canko:2020ylt,Abreu:2021smk,Abreu:2023rco} yields purely rational-number coefficients multiplying a small kinematic basis $\{1,B,B_1\}$, manifesting maximal transcendentality and matching the planar result~\cite{Guo:2024bsd}. Evaluating the UT basis in terms of pentagon functions~\cite{Chicherin:2021dyp,Abreu:2023rco}, we verify exact infrared-pole cancellation and extract the full-color finite remainder in functional and symbol form. The parity-odd factor $B$ cancels completely in the full-color remainder.

Our result provides the first full-color CK-dual integrand construction and analytic evaluation of a four-point form factor, i.e., a two-loop five-point process with one massive external leg. As non-trivial physical checks and applications, we verify the consistency of double-collinear and single-soft factorization and extract new non-planar two-loop infrared data: the subleading-color triple-collinear splitting function and double-soft current. By the principle of maximal transcendentality~\cite{Kotikov:2004er,Brandhuber:2012vm,Gehrmann:2011aa,Guo:2022pdw}, our results are also expected to shed light on the non-planar structure of two-loop Higgs-plus-four-parton amplitudes in QCD.

\section{CK-dual integrand construction}

\begin{figure}[t]
\centering
\includegraphics[height=1.4cm]{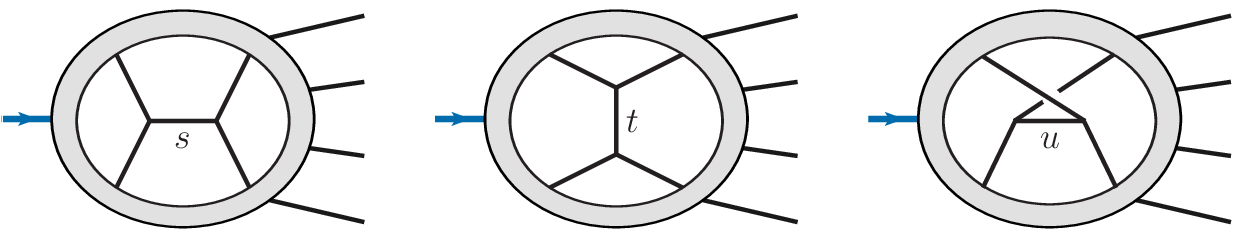}
\caption{Color-kinematics duality at loop level: the $s$-, $t$- and $u$-channel graphs are related by Jacobi identities.}
\label{fig:loopBCJ}
\end{figure}

We construct the CK-dual one- and two-loop integrands for the four-point form factor in two steps (see, e.g.,~\cite{Bern:2012uf, Boels:2012ew}): (i) build an ansatz in which all kinematic numerators are generated from a small basis of master numerators via dual Jacobi relations; and (ii) determine the unknown coefficients by imposing graph symmetries and generalized-unitarity cuts. In this trivalent representation, the $L$-loop full-color four-point form factor takes the form
\begin{equation}
\mathbf{F}_{4}^{(L)}=
 \sum_{\sigma\in S_4}\sum_{i=1}^{n}
 \int\!\prod_{r=1}^{L} \frac{d^D\ell_r}{(2\pi)^D}\,
\frac{1}{S_i} \frac{C_iN_i}{\prod_{\alpha_i} P_{\alpha_i}^{\,2}}\, .
 \label{eq:gauge-integrand}
\end{equation}
Here $S_i$ is the symmetry factor of the $i$-th trivalent graph, while $C_i$ and $N_i$ denote its color factor and kinematic numerator, respectively. For each internal propagator, the color Jacobi identity $C_s=C_t+C_u$ is mirrored by a dual kinematic Jacobi relation $N_s=N_t+N_u$ among the corresponding numerators, as illustrated in Fig.~\ref{fig:loopBCJ}.

\subsection{Warm-up: One-loop integrand construction}

\begin{figure}[t]
\centering
\includegraphics[width=\columnwidth]{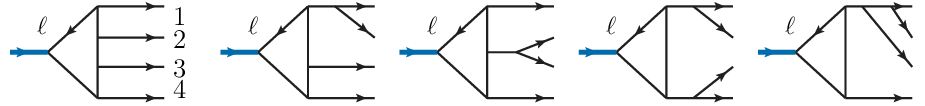}
\caption{The five trivalent topologies contributing to the one-loop four-point form factor.}
\label{fig:1l-topos}
\end{figure}

At one loop, there are five distinct trivalent topologies for the four-point form factor, as shown in Fig.~\ref{fig:1l-topos}. Dual Jacobi relations are generated for each propagator of a topology. 
The first topology can be chosen as the master topology, whose numerator $N_1$ can determine the remaining ones via:
\begin{equation}
\label{eq:1l-ck}
\begin{aligned}
N_{2}(1,2,3,4;\ell) & =
N_{1}(1,2,3,4;\ell)-N_{1}(2,1,3,4;\ell)\,,\\
N_{3}(1,2,3,4;\ell) & =
N_{1}(1,2,3,4;\ell)-N_{1}(1,3,2,4;\ell)\,,\\
N_{4}(1,2,3,4;\ell) & =
N_{2}(1,2,3,4;\ell)-N_{2}(1,2,4,3;\ell)\,,\\
N_{5}(1,2,3,4;\ell) & =
N_{2}(1,2,3,4;\ell)-N_{3}(3,1,2,4;\ell)\,,
\end{aligned}
\end{equation}
where $1,\dots,4$ denote the massless external momenta $p_1,\dots,p_4$ and $\ell$ is the loop momentum. Thus, it suffices to construct an ansatz for $N_1$.

Unlike the Sudakov and three-point form factors \cite{Boels:2012ew,Yang:2016ear,Lin:2021kht,Lin:2021lqo}, the full-color four-point form factor has a non-trivial spinor-helicity structure. Drawing insight from the CK-dual construction of the five-point amplitude \cite{Carrasco:2011mn,Cai:2026cay},
we introduce the following $\beta$ factor:
\begin{equation}
\label{eq:1l-beta-def}
\beta_{1234}\equiv\delta^{(8)}(Q)\,\frac{[12][23][34][41]}{{\rm Gram}(p_{1},p_{2},p_{3},p_{4})}\,,
\end{equation}
where $Q^{\alpha A}=\sum_{i} \lambda_i^\alpha \eta_i^{A}$ and ${\rm Gram}(p_{1},p_{2},p_{3},p_{4})$ is the Gram determinant. This $\beta$ prefactor satisfies several symmetries:
\begin{equation}
\label{eq:1l-beta-cyclic}
\beta_{1234}=\beta_{2341}\,, \quad \beta_{1234}=\beta_{3214}\,, \quad \beta_{1234}=\beta_{1432} \,,
\end{equation}
which implies that only three $\beta$ factors are linearly
independent and we can choose the basis as $\beta_{1234},\beta_{1243},\beta_{1324}$. 

We build the ansatz for $N_1$ from this $\beta$ basis dressed with Lorentz contractions of external and loop momenta (with mass dimension $6$). In $\mathcal{N}=4$ SYM, UV power counting restricts a one-loop $n$-point subgraph with one massive leg to carry at most $n-3$ powers of $(p_i\cdot\ell)$~\cite{Bern:2012uf, Boels:2012ew}, while allowing inverse-propagator factors $\ell^2$ that generate contact terms in the CK-dual representation. Consequently, the candidate monomials fall into six classes:
\[\{s^{3}, \, s^{2}(p_{i}\cdot \ell), \, s(p_{i}\cdot \ell)^{2}, \,
 s^{2}\ell^{2}, \, s\ell^{2}(p_{i}\cdot \ell), \, s(\ell^{2})^{2}\} \times \beta\,.\]
Furthermore, linear identities relate products of $\beta$ and Mandelstam invariants, such as
\begin{equation}
\label{eq:1l-beta-mandelstam}
\beta_{1243}\,s_{14}s_{23}+\beta_{1234}\,s_{13}s_{24}+\beta_{1324}\,s_{12}s_{34}=0\,.
\end{equation}
Accounting for all such redundancies leaves $742$ independent monomials in the ansatz for $N_1$.

\begin{figure}[t]
\centering
\includegraphics[width=\columnwidth]{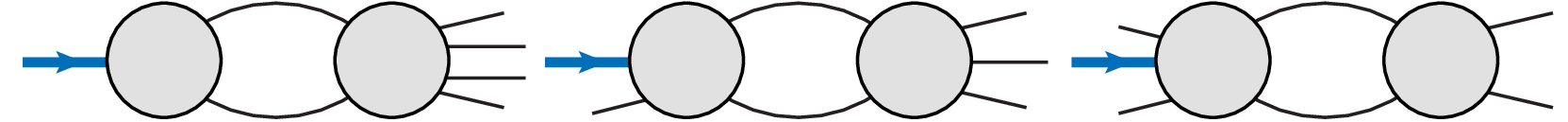}
\caption{The three four-dimensional unitarity cuts used to solve the ansatz.}
\label{fig:1l-cuts}
\end{figure}

Imposing the automorphism symmetries of the topologies reduces the number of free parameters to $398$, and the three four-dimensional unitarity cuts in Fig.~\ref{fig:1l-cuts} further fix all but $98$ parameters. Upon projecting the integrand onto an independent integral basis, all $98$ remaining parameters cancel out, confirming that they correspond purely to generalized-gauge degrees of freedom. Extracting the single-trace component, we find complete agreement with the one-loop integrand of~\cite{Guo:2024bsd}. Notably, our CK-dual solution contains no explicit $\mu$-terms, yet reproduces the $\mu$-term-dependent integrand of~\cite{Guo:2024bsd}—a feature that persists at two loops.

\subsection{Two-loop integrand construction}

\begin{figure*}[t]
\centering
\includegraphics[width=\textwidth]{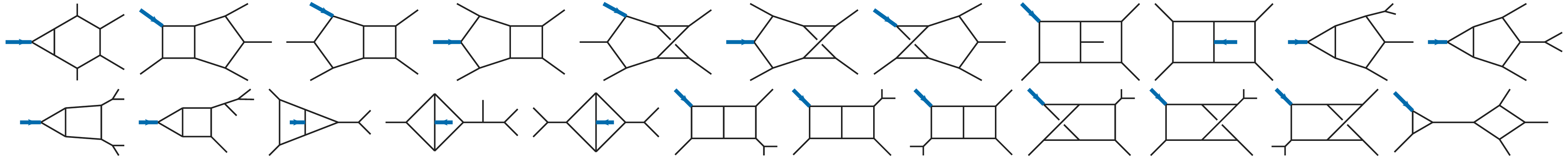}
\caption{The 23 trivalent topologies for the two-loop four-point form factor.}
\label{fig:2l-topos}
\end{figure*}

The two-loop CK-dual construction follows the same strategy, so we highlight only new features. There are $23$ trivalent topologies, shown in Fig.~\ref{fig:2l-topos}. The numerator of the last graph in the second row vanishes under symmetry and cut constraints and is omitted in practice. The color factors of the two graphs proceeding it in the second row vanish and do not contribute to the final form factor, yet they are needed to solve the Jacobi relations and must be kept in the CK-dual construction. 
After solving all CK relations, we choose the first three graphs as the master topologies.

As at one loop, the ansatz for the three numerators is built from the $\beta$ factor and the Lorentz contractions between external and loop momenta. 
At two loops, the mass dimension is 8, so the monomials are
products of four Lorentz contractions and $\beta$. 
After UV power counting and linear relations among monomials,
the numbers of independent monomials in the 
three numerators are 2097, 3527 and 2097, respectively, giving 7721 undetermined coefficients in total.

\begin{figure}[t]
\centering
\includegraphics[width=\columnwidth]{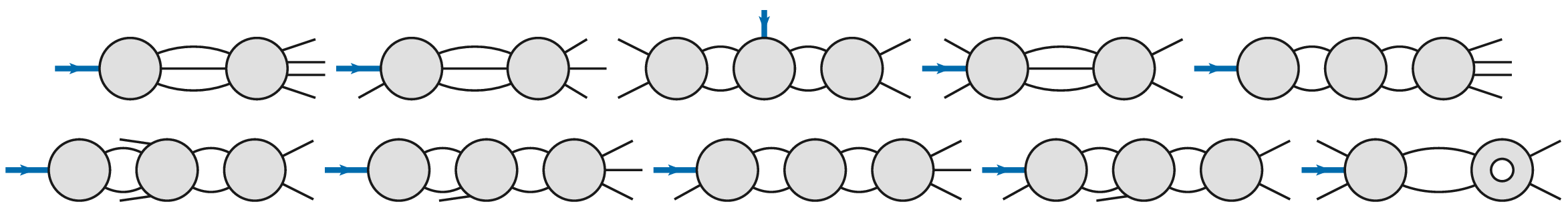}
\caption{The 10 unitarity cuts used to solve the two-loop ansatz. 
The last one carries $D$-dimensional information.
}
\label{fig:2l-cuts}
\end{figure}

Graph symmetry constraints leave 1363
undetermined coefficients. We then impose the unitarity-cut constraints, as shown in
Fig.~\ref{fig:2l-cuts}, which leave 256 final free parameters. 
The first three cuts already suffice to fix the result, while the
remaining seven serve as consistency checks.
The last cut contains a four-point one-loop sub-amplitude and therefore retains part of the
$D$-dimensional loop-momentum information, whereas the other cuts are four-dimensional.

After projecting onto a basis of independent integrals (see e.g., \cite{Lin:2021qol}), we verify that all 256 generalized-gauge parameters cancel, and that our planar integrand is equivalent to the two-loop integrand in \cite{Guo:2024bsd} in $D$ dimensions. It is worth stressing that the integrand in \cite{Guo:2024bsd} contains $\mu$-terms
and traces of $D$-dimensional $\Gamma$ matrices, which contain information beyond
four dimensions, while our integrand does not contain them explicitly. The CK-dual construction built from the $\beta$-factor ansatz naturally encodes the $D$-dimensional information of the four-point form factor. This is a strong check of our result.

We provide both one- and two-loop CK-dual integrands in the ancillary files.

\section{Color structure}

To facilitate the discussion later, we consider color decomposition of the four-point form factor. 
The one-loop form factor can be written as the sum of single-trace and double-trace
parts
\begin{equation}\label{eq:1l-color}
\begin{aligned}
\mathbf{F}^{(1)}_{4}={}&N_{c}\Big[\mathcal{F}^{(1,0)}_{4}(1234)\,{\rm tr}(1234)
+5\text{ perms}\Big]\\
&+\Big[\mathcal{F}^{(1,1)}_{4}(12;34)\,{\rm tr}(12)\,{\rm tr}(34)+2\text{ perms}\Big]\,.
\end{aligned}
\end{equation}
Here we follow the notation of \cite{Naculich:2011ep} to label contributions of
different $N_{c}$ powers. The one-loop double-trace part
can be expanded in terms of single-trace parts as \cite{Bern:1994zx}:
\begin{equation}\label{eq:1l-dt}
\mathcal{F}^{(1,1)}_{4}(12;34)=2\!\Big[\mathcal{F}^{(1,0)}_{4}(1234)
+\mathcal{F}^{(1,0)}_{4}(1243)\\
+\mathcal{F}^{(1,0)}_{4}(1324)\Big]\,.
\end{equation}
Thus the three double-trace parts are equal.

At two loops, the form factor decomposes into planar, subleading $N_{c}$
single-trace, and double-trace parts:
\begin{align}\label{eq:2l-traces}
\mathbf{F}^{(2)}_{4}={}&N_{c}^{2}\Big[\mathcal{F}^{(2,0)}_{4}(1234)\,{\rm tr}(1234)
+5\text{ perms}\Big]\\
&+\Big[\mathcal{F}^{(2,2)}_{4}(1234)\,{\rm tr}(1234)+5\text{ perms}\Big] \nonumber\\
&+N_{c}\Big[\mathcal{F}^{(2,1)}_{4}(12;34)\,{\rm tr}(12)\,{\rm tr}(34)+2\text{ perms}\Big]\,. \nonumber
\end{align}
The relations between non-planar and the planar parts of two-loop four-point amplitudes~\cite{Naculich:2011ep} also apply to the four-point form factor. First,
the subleading $N_{c}$ single-trace contribution satisfies
\begin{equation}\label{eq:2l-stsub-sum}
\mathcal{F}^{(2,2)}_{4}(1234)+\mathcal{F}^{(2,2)}_{4}(1324)+\mathcal{F}^{(2,2)}_{4}(1243)=0\,.
\end{equation}
Moreover, the double-trace part can be expressed through the single-trace parts,
\begin{equation}\label{eq:2l-double-trace}
\begin{aligned}
& \Sigma=\mathcal{F}^{(2,0)}_{4}(1234)+\mathcal{F}^{(2,0)}_{4}(1324)+\mathcal{F}^{(2,0)}_{4}(1243)\,,\\
& \mathcal{F}^{(2,1)}_{4}(12;34)=2\Sigma-\mathcal{F}^{(2,2)}_{4}(1324)\,.
\end{aligned}
\end{equation}

\section{Integral reduction}

To evaluate the form factor, we perform IBP reduction \cite{Chetyrkin:1981qh, Tkachov:1981wb} using Kira and Fermat~\cite{Maierhofer:2017gsa, Klappert:2020nbg, Lewis} onto the uniform-transcendentality (UT) basis $I^{(2),{\rm UT}}_{i}$
of \cite{Abreu:2020jxa, Canko:2020ylt, Abreu:2021smk, Abreu:2023rco}. 
After reduction, we obtain $\mathcal{F}^{(2,0)}_{4}(1234)$, $\mathcal{F}^{(2,2)}_{4}(1234)$ and $\mathcal{F}^{(2,1)}_{4}(12;34)$ and find that they satisfy Eqs.~\eqref{eq:2l-stsub-sum} and \eqref{eq:2l-double-trace}. The planar part agrees with the
result in \cite{Guo:2024bsd}, and the double-trace part follows from the
single-trace part, so below we focus on the subleading $N_{c}$ single-trace
contribution.

We extract an overall tree-level factor,
\begin{equation}\label{eq:2l-extract}
\mathcal{F}^{(2,2)}_{4}(1234)=\mathcal{F}^{(0)}_{4}(1234)\,\mathcal{I}^{(2,2)}(1234)\,,
\end{equation}
where
\begin{equation}\label{eq:tree-ff}
\mathcal{F}^{(0)}_{4}(1234)=\frac{\delta^{(8)}(Q)}
{\langle 12\rangle\langle 23\rangle\langle 34\rangle\langle 41\rangle}\,.
\end{equation}
We find that $\mathcal{I}^{(2,2)}(1234)$ can be expanded in the basis $\{1,B,B_{1}\}$ as
\begin{equation}\label{eq:2l-expansion}
\mathcal{I}^{(2,2)}(1234)=\sum_{i=1}^{1317}\Big(a_{i}+b_{i}\,B+c_{i}\,B_{1}\Big)
I^{(2),{\rm UT}}_{i}\,,
\end{equation}
where the factor $B$ is defined as
\begin{equation}\label{eq:B-factor}
B=\frac{s_{12}s_{34}+s_{14}s_{23}-s_{13}s_{24}}{4i\varepsilon(1234)}\,.
\end{equation}
and $B_{1}$ is a ratio of two tree-level factors,
\begin{equation}\label{eq:B1-factor}
B_{1}=\frac{\langle 12\rangle\langle 34\rangle}{\langle 13\rangle\langle 24\rangle}
=-\frac{\mathcal{F}^{(0)}_{4}(1324)}{\mathcal{F}^{(0)}_{4}(1234)}\,.
\end{equation}
A non-trivial consistency check is that the expansion coefficients $\{a_{i},b_{i},c_{i}\}$ are all simple rational numbers as expected from the uniform transcendentality of the half-BPS form facotr.
The factor $B_{1}$ also appears in the two-loop four-point $\mathrm{tr}(\phi^{3})$ form factor~\cite{Guo:2021bym}. 
The UT-master expansion results are provided in the ancillary files.

\section{Finite hard function}
\label{sec:IRsubtraction}

In this section we consider the form factor at function level, first discussing the IR pole subtraction and then the property of the finite hard function.

In dimensional reduction ($D=4-2\epsilon$), the form factor is expanded as Laurent expansion in $\epsilon$ which is straightforward to obtain by expressing the UT master integrals as one-mass pentagon functions \cite{Chicherin:2021dyp,Abreu:2023rco}. 
Since ${\cal T}$ is a half-BPS operator, loop corrections of the form factor are free of ultraviolet divergences and all divergences are purely infrared (soft and collinear). 
The full-color two-loop IR divergences of amplitudes and form factors is well understood \cite{Catani:1998bh, Aybat:2006mz, Becher:2009cu, Gardi:2009qi, Becher:2009qa}. 
Here we use the dipole subtraction scheme with following factorization property \cite{Becher:2009cu, Gardi:2009qi}
\begin{equation}
\label{eq:factZ}
\mathbf{F}_4 = {\mathbf{Z}}\,\mathbf{F}_{4,\mathrm{hard}} \,,
\end{equation}
where $\mathbf{Z}$ factor captures all divergences and $\mathbf{F}_{4,\mathrm{hard}}$ is finite as $\epsilon \to 0$.
Introducing the ${\cal N}=4$ version of one-loop and two-loop IR operators:
\begin{align}
\label{eq:Z1W2}
\mathbf{Z}_1 &= \frac{2\mathbf{D}_0}{\epsilon^2} + \frac{2\mathbf{D}_1}{\epsilon}  \,, \nonumber\\
\mathbf{W}_2 &= -\frac{N_c\zeta_2 \mathbf{D}_0}{\epsilon^2} - \frac{N_c\zeta_3 \mathbf{D}_0 + 2N_c\zeta_2\mathbf{D}_1}{\epsilon} \,,
\end{align}
where
\begin{equation}
\mathbf{D}_0 = \sum_{1\leq i<j\leq 4} \hat{\mathbf{T}}_i\cdot \hat{\mathbf{T}}_j  \,, \quad 
\mathbf{D}_1 = \sum_{1\leq i<j\leq 4} \hat{\mathbf{T}}_i\cdot \hat{\mathbf{T}}_j \log\frac{\mu^2}{-s_{ij}} \,,
\end{equation}
and $\hat{\mathbf{T}}_j^b$ are operators acting on color factors as $\hat{\mathbf{T}}_j^b T^{a_i} = - \textrm{i} f^{b a_i c_i} T^{c_i} \delta_j^i$,
the two-loop finite part is then extracted as
\begin{align}
\label{eq:F2MS}
\mathbf{F}_{4,\mathrm{fin}}^{(2)} &= \Big[ \mathbf{F}_4^{(2)} - \mathbf{Z}_1 \mathbf{F}_4^{(1)} + (\frac{1}{2}\mathbf{Z}_1^2 - \mathbf{W}_2)\mathbf{F}_4^{(0)} \Big]_{\epsilon^0} \nonumber\\
&= \big[\mathbf{F}_4^{(2)}\big]_{0} + 4N_c \big[\mathbf{F}_4^{(1)}\big]_2 - 2\mathbf{D}_1 \big[\mathbf{F}_4^{(1)}\big]_1 \,,
\end{align}
where $\mathbf{F}^{(l)} = \sum_{r} \epsilon^r [\mathbf{F}^{(l)}]_r$.
We have checked that IR singularities consistently cancel for all nine color-trace components.
Below we consider the finite part and follow the same color decomposition \eqref{eq:2l-traces}.

As a first check, the leading-color planar part $\mathcal{R}_{4}^{(2,0)} = \mathcal{F}_{4,\mathrm{fin}}^{(2,0)}/ \mathcal{F}_{4}^{(0)}$ reproduces the known BDS remainder function $\mathcal{R}_{4,\mathrm{BDS}}^{(2)}$~\cite{Guo:2024bsd} via
\begin{equation}
\label{eq:MSvsBDS}
\mathcal{R}_{4}^{(2,0)} - \mathcal{R}_{4,\mathrm{BDS}}^{(2)} = \frac{1}{2} h_1^2 - 2\zeta_2 h_1 + \zeta_3 \sum_{i=1}^4 L_{i i+1} + 2\zeta_4 \,,
\end{equation}
where $h_1 = [\mathcal{I}_{4}^{(1,0)}]_{0}$ and $L_{ij}=\log\frac{\mu^2}{-s_{ij}}$. 

For the subleading single-trace term $\mathcal{F}_{4}^{(2,2)} = \mathcal{F}_4^{(0)} \mathcal{I}_{4}^{(2,2)}$, color analysis shows that $\mathcal{F}_{4}^{(2,2)}$ is free of $\epsilon^{-4}$ and $\epsilon^{-3}$ poles, and the infrared counterterms only originate from the dipole operator fusing one-loop double traces into single traces. The finite remainder reduces to
\begin{equation}
\label{eq:Rsubleading}
\mathcal{R}_{(1234)}^{(2,2)} 
= \big[\mathcal I_{(1234)}^{(2,2)}\big]_0-\frac{\big[\mathcal F^{(1,1)}_{(12;34)}\big]_1}{\mathcal F_{(1234)}^{(0)}}\,\log\frac{s_{12}s_{23}s_{34}s_{41}}{(s_{13}s_{24})^2} \,.
\end{equation}
The finite function for the double-trace part can be obtained using \eqref{eq:2l-double-trace}.

An important observation is that the full-color finite remainder is free of the $B$ factor, even though it appears in the master expansion \eqref{eq:2l-expansion}.
In the planar case, the cancellation of $B$ factor is crucial to preserve the duality between the form factor and the periodic Wilson loop \cite{Guo:2024bsd}.

The symbol \cite{Goncharov:2010jf} of the remainder can be obtained using the Chen iterated path integral representation \cite{Chen:1977oja}. We check the planar symbol matches the known result \cite{Dixon:2022xqh}.
The non-planar part is more complicated; in particular, 
the hidden directional dual conformal symmetry observed in the planar lightlike limit $q^2\to 0$~\cite{Guo:2022qgv} is broken by non-planar corrections.
We provide both pentagon-function and symbol results in the ancillary files.

\section{Collinear and soft limits}
\label{sec:limits}

As applications of the result, we consider following collinear and soft limits, which provide non-trivial consistency checks and also new results:
\paragraph{Double- and triple-collinear limits:}
\begin{align}
\label{eq:double-collinear}
\mathbf{F}_4^{(2)} & \xlongrightarrow{p_1 \parallel p_2} \sum_{l=0}^2 \mathbf{Sp}^{(l)} (p_1,p_2,P) \mathbf{F}^{(2-l)}_3(-P,p_3,p_4) \,, \\
\label{eq:triple-collinear}
\mathbf{F}_4^{(2)} & \xlongrightarrow{p_1 \parallel p_2 \parallel p_3} \sum_{l=0}^2 \mathbf{Sp}^{(l)}(p_1, p_2, p_3,P) \mathbf{F}^{(2-l)}_2(-P, p_4) \,.
\end{align}
\paragraph{Single- and double-soft limits:}
\begin{align}
\label{eq:single-soft}
\mathbf{F}_4^{(2)} & \xlongrightarrow{p_1 \rightarrow 0} \sum_{l=0}^2 \mathbf{J}^{(l)}(p_1) \mathbf{F}^{(2-l)}_3(p_2, p_3, p_4) \,, \\
\label{eq:double-soft}
\mathbf{F}_4^{(2)} & \xlongrightarrow{p_1,p_2 \rightarrow 0} \sum_{l=0}^2 \mathbf{J}^{(l)}(p_1,p_2) \mathbf{F}^{(2-l)}_2(p_3, p_4) \,.
\end{align}
Since the lower two- and three-point form factors have only leading-color planar parts up to two loops \cite{Brandhuber:2012vm}, non-planar corrections can arise only in the splitting or soft factors. 

For the double-collinear limit, the two-particle splitting amplitude has no subleading-color corrections through two loops \cite{Bern:2004cz}, which requires that the nonplanar corrections remain finite in the limit. 
We find that the subleading single-trace remainder satisfies
\begin{equation}
    \label{eq:collinearzero}
    \lim_{s_{12}\rightarrow0}  \mathcal{R}_{{(1234)}}^{(2,2)} = \sqrt{s_{12}} \Big[ 4 \log^4 s_{12} + {\cal O}( \log^3 s_{12}) \Big] \rightarrow0  \,.
\end{equation}
The $\sqrt{s_{12}}$ factor is due to the tree factor $\mathcal{F}_{\mathrm{fin},(1234)}^{(2,2)} \propto \mathcal{R}_{(1234)}^{(2,2)}/{\langle1 2 \rangle}$. 
For the single-soft limit, we check that our result has consistent behavior using the full-color soft factor \cite{Dixon:2019lnw}.

The triple-collinear and double-soft limits give new non-planar results, as only planar results were previously known for the two-loop three-particle splitting \cite{Bern:2008ap} and double-soft factors \cite{Carrolo:2025agz}.
For the subleading single-trace term, since $\mathbf{F}_2^{(l)}\propto N_c^{l}$, only the $l=2$ terms of \eqref{eq:triple-collinear} and \eqref{eq:double-soft} contribute at order $N_c^0$, yielding
\begin{align}
\label{eq:np-triple-collinear}
\mathcal{I}_{(1234)}^{(2,2)} & \xlongrightarrow{p_1 \parallel p_2 \parallel p_3} \frac{\mathrm{Sp}^{(2,2)}}{\mathrm{Sp}^{(0)}}
= \Big(\frac{\mu^2}{-s_{123}}\Big)^{2\epsilon} \sum_{r} r\, \mathcal{C}_r \,, \\
\label{eq:np-double-soft}
\mathcal{I}_{(1234)}^{(2,2)} & \xlongrightarrow{\ p_1,p_2 \rightarrow 0\ } \frac{\mathrm{J}^{(2,2)}}{\mathrm{J}^{(0)}}
= \Big(\frac{\mu^2}{-s_{12}}\Big)^{2\epsilon} \sum_{r} r\, \mathcal{S}_r \,.
\end{align}
where $r\in\{1,B_1,B\}$ as in \eqref{eq:2l-expansion}.
Here $\mathrm{Sp}^{(2,2)}$ and $\mathrm{J}^{(2,2)}$ are the $N_c^0\,{\rm tr}(1234)$ components of $\mathbf{Sp}^{(2)}$ and $\mathbf{J}^{(2)}$, normalized by their tree-level factors (in the MHV sector, i.e.~for same-helicity soft gluons emitted from two hard legs).

Through ${\cal O}(\epsilon^0)$ the $B$ terms stem solely from the ${\cal O}(\epsilon)$ part of the one-loop double-trace factors and cancel after IR subtraction, cf.~\eqref{eq:Rsubleading}.
The genuinely new two-loop information resides in the weight-four functions $\mathcal{C}_{1}$, $\mathcal{C}_{B_1}$, $\mathcal{S}_{1}$ and $\mathcal{S}_{B_1}$ at ${\cal O}(\epsilon^0)$, whose symbol are provided in the ancillary files.
The double-trace part follows from \eqref{eq:2l-double-trace}.

We comment that the factors $B_1$ and $B$ stay finite, since both limits leave the cross ratios $u=\frac{s_{12}s_{34}}{s_{13}s_{24}}=|B_1|^2$ and $v=\frac{s_{14}s_{23}}{s_{13}s_{24}}=|1-B_1|^2$ generic.
The coefficients $\mathcal{C}_r$ and $\mathcal{S}_r$ are pure functions of uniform transcendentality starting at $\epsilon^{-2}$: $\mathcal{C}_r$ depend only on the momentum fractions $z_i$ ($p_i\to z_i P$) and $s_{ij}/s_{123}$, and $\mathcal{S}_r$ only on $u$, $v$ and $s_{13}/s_{23}$.
As a check, for $p_1\ll p_2$ the double-soft result reduces to the tripole term of the two-loop single-soft current \cite{Dixon:2019lnw}.

\section{Summary}

In this Letter, we construct the first full-color two-loop form factor via CK duality based on the $\beta$-factor ansatz \eqref{eq:1l-beta-def}.
After IBP reduction, we obtain the UT master expansion, check the IR poles cancel and extract the finite part. 
We perform nontrivial checks in the double-collinear and single-soft limits, and obtain new results for the non-planar triple splitting function and the double-soft current.
Although computed for ${\cal N}=4$ SYM, these results are expect to capture the maximally transcendental part of QCD Higgs-plus-four-parton amplitudes for certain helicity sectors \cite{Guo:2022pdw}.
It would be interesting to generalize the present construction to the non-MHV four-point case, and to higher loops, which we leave for future work.

\vskip .3cm

{\it Acknowledgments.}
We thank Xinyue Li, Lei Wang and Guorui Zhu for discussions.
In particular, we thank Song He and Xuhang Jiang for sharing their progress based on a different bootstrap approach, cross-checking the remainder results and coordinating the submission of the paper to arXiv.
This work was largely completed without using AI, which at the final stage assisted in generating figures, checking and exporting data files and refining the manuscript.
This work is supported by the National Natural Science Foundation of
China (Grants No.~12425504, 12447101, 12247103) and by the Chinese
Academy of Sciences (Grant No.~YSBR-101).
We also acknowledge the support of the HPC Cluster of ITP-CAS.

\bibliographystyle{apsrev4-2}

\bibliography{formfactor}

\end{document}